\documentclass[useAMS,usenatbib]{mn2e}
\usepackage{epsfig}
\usepackage{graphicx}
\usepackage{epstopdf}
\DeclareGraphicsExtensions{.ps}
\usepackage{mathptmx}
\usepackage{natbib}
\usepackage{amssymb}
\usepackage{amsmath}

\newif\ifAMStwofonts
\AMStwofontstrue
\newcommand{\be}{\begin{equation}}
\newcommand{\ee}{\end{equation}}
\newcommand{\bes}{\begin{equation*}}
\newcommand{\ees}{\end{equation*}}
\newcommand{\ba}{\begin{eqnarray}}
\newcommand{\ea}{\end{eqnarray}}
\newcommand{\bas}{\begin{eqnarray*}}
\newcommand{\eas}{\end{eqnarray*}}
\newcommand{\brr}{\begin{array}}
\newcommand{\err}{\end{array}}
\newcommand{\bc}{\begin{center}}
\newcommand{\ec}{\end{center}}
\newcommand{\bit}{\begin{itemize}}
\newcommand{\eit}{\end{itemize}}

\newcommand{\gz}{\mbox{$\rm{GHz}$ } }

\newcommand{\mincir}{\raise
  -2.truept\hbox{\rlap{\hbox{$\sim$}}\raise5.truept \hbox{$<$}\ }}
\newcommand{\magcir}{\raise
  -2.truept\hbox{\rlap{\hbox{$\sim$}}\raise5.truept \hbox{$>$}\ }}
\newcommand{\siml}{\raise
  -2.truept\hbox{\rlap{\hbox{$\sim$}}\raise5.truept \hbox{$<$}\ }}
\newcommand{\simg}{\raise
  -2.truept\hbox{\rlap{\hbox{$\sim$}}\raise5.truept \hbox{$>$}\ }}

\title[SPT Constraints on the CMB Temperature
Evolution] {Constraints on the CMB Temperature Evolution using
Multi-Band Measurements of the Sunyaev Zel'dovich Effect with the
South Pole Telescope} 

\newcommand{\Munich}{$^{1}$}
\newcommand{\ExcellenceCluster}{$^{2}$}
\newcommand{\MPE}{$^{3}$}
\newcommand{\UChicago}{$^{4}$}
\newcommand{\CfA}{$^{5}$}
\newcommand{\Harvard}{$^{6}$}
\newcommand{\FNAL}{$^{7}$}
\newcommand{\KICPChicago}{$^{8}$}
\newcommand{\EFIChicago}{$^{9}$}
\newcommand{\PhysicsUChicago}{$^{10}$}
\newcommand{\ANL}{$^{11}$}
\newcommand{\Miss}{$^{12}$}
\newcommand{\AAUChicago}{$^{13}$}
\newcommand{\NIST}{$^{14}$}
\newcommand{\PUC}{$^{15}$}
\newcommand{\McGill}{$^{16}$}
\newcommand{\illast}{$^{17}$}
\newcommand{\illphy}{$^{18}$}
\newcommand{\Berkeley}{$^{19}$}
\newcommand{\UFlorida}{$^{20}$}
\newcommand{\Colorado}{$^{21}$}
\newcommand{\LBNL}{$^{22}$}
\newcommand{\Caltech}{$^{23}$}
\newcommand{\Arizona}{$^{24}$}
\newcommand{\MIT}{$^{25}$}
\newcommand{\Michigan}{$^{26}$}
\newcommand{\CaseWestern}{$^{27}$}
\newcommand{\Minnesota}{$^{28}$}
\newcommand{\STScI}{$^{29}$}
\newcommand{\SAIC}{$^{30}$}
\newcommand{\Dunlap}{$^{31}$}
\newcommand{\Toronto}{$^{32}$}

\author[A. Saro, et al.]
{A.~Saro\Munich$^,$\ExcellenceCluster,
J.~Liu\Munich$^,$\ExcellenceCluster,
J.~J.~Mohr\Munich$^,$\ExcellenceCluster$^,$\MPE,
K.~A.~Aird\UChicago,
M.~L.~N.~Ashby\CfA,
M.~Bayliss\CfA$^,$\Harvard,
\newauthor 
B.~A.~Benson\FNAL$^,$\KICPChicago$^,$\EFIChicago,
L.~E.~Bleem\KICPChicago$^,$\PhysicsUChicago$^,$\ANL,
S.~Bocquet\Munich$^,$\ExcellenceCluster,
M.~Brodwin\Miss,
\newauthor 
J.~E.~Carlstrom\KICPChicago$^,$\EFIChicago$^,$\PhysicsUChicago$^,$\ANL$^,$\AAUChicago,
C.~L.~Chang\KICPChicago$^,$\EFIChicago$^,$\ANL,
I.~Chiu\Munich$^,$\ExcellenceCluster,
H.~M. Cho\NIST,
A.~Clocchiatti\PUC,
\newauthor 
T.~M.~Crawford\KICPChicago$^,$\AAUChicago,
A.~T.~Crites\KICPChicago$^,$\AAUChicago,
T.~de~Haan\McGill,
S.~Desai\Munich$^,$\ExcellenceCluster,
J.~P.~Dietrich\Munich$^,$\ExcellenceCluster,
\newauthor 
M.~A.~Dobbs\McGill,
K.~Dolag\Munich$^,$\ExcellenceCluster,
J.~P.~Dudley\McGill,
R.~J.~Foley\illast$^,$\illphy,
D.~Gangkofner\Munich$^,$\ExcellenceCluster,
\newauthor
E.~M.~George\Berkeley,
M.~D.~Gladders\KICPChicago$^,$\AAUChicago,
A.~H.~Gonzalez\UFlorida,
N.~W.~Halverson\Colorado,
\newauthor
C.~Hennig\Munich$^,$\ExcellenceCluster,
W.~L.~Holzapfel\Berkeley,
J.~D.~Hrubes\UChicago,
C.~Jones\CfA,
R.~Keisler\KICPChicago$^,$\PhysicsUChicago,
A.~T.~Lee\Berkeley$^,$\LBNL,
\newauthor
E.~M.~Leitch\KICPChicago$^,$\AAUChicago,
M.~Lueker\Berkeley$^,$\Caltech,
D.~Luong-Van\UChicago,
A.~Mantz\KICPChicago,
D.~P.~Marrone\Arizona,
\newauthor 
M.~McDonald\MIT,
J.~J.~McMahon\Michigan,
J.~Mehl\KICPChicago$^,$\AAUChicago,
S.~S.~Meyer\KICPChicago$^,$\EFIChicago$^,$\PhysicsUChicago$^,$\AAUChicago,
L.~Mocanu\KICPChicago$^,$\AAUChicago,
\newauthor
T.~E.~Montroy\CaseWestern,
S.~S.~Murray\CfA,
D.~Nurgaliev\Harvard,
S.~Padin\KICPChicago$^,$\AAUChicago$^,$\Caltech,
A.~Patej\Harvard,
C.~Pryke\Minnesota,
\newauthor
C.~L.~Reichardt\Berkeley,
A.~Rest\STScI,
J.~Ruel\Harvard,
J.~E.~Ruhl\CaseWestern,
B.~R.~Saliwanchik\CaseWestern,
J.~T.~Sayre\CaseWestern,
\newauthor
K.~K.~Schaffer\KICPChicago$^,$\EFIChicago$^,$\SAIC,
E.~Shirokoff\Berkeley$^,$\Caltech,
H.~G.~Spieler\LBNL,
B.~Stalder\CfA,
Z.~Staniszewski\CaseWestern,
\newauthor
A.~A.~Stark\CfA,
K.~Story\KICPChicago$^,$\PhysicsUChicago,
A.~van~Engelen\McGill,
K.~Vanderlinde\Dunlap$^,$\Toronto,
J.~D.~Vieira\illast$^,$\Caltech,
\newauthor
A. Vikhlinin\CfA,
R.~Williamson\KICPChicago$^,$\AAUChicago,
O.~Zahn\Berkeley, 
A.~Zenteno\Munich$^,$\ExcellenceCluster 
\\
\\
\Munich Department of Physics, Ludwig-Maximilians-Universit\"{a}t, Scheinerstr.\ 1, 81679 M\"{u}nchen, Germany \\
\ExcellenceCluster Excellence Cluster Universe, Boltzmannstr.\ 2, 85748 Garching, Germany \\
\MPE Max-Planck-Institut f\"{u}r extraterrestrische Physik, Giessenbachstr.\ 85748 Garching, Germany \\
\UChicago University of Chicago, 5640 South Ellis Avenue, Chicago, IL 60637 \\
\CfA Harvard-Smithsonian Center for Astrophysics, 60 Garden Street, Cambridge, MA 02138 \\
\Harvard Department of Physics, Harvard University, 17 Oxford Street, Cambridge, MA 02138 \\
\FNAL Center for Particle Astrophysics, Fermi National Accelerator Laboratory, Batavia, IL, USA 60510 \\
\KICPChicago Kavli Institute for Cosmological Physics, University of Chicago, 5640 South Ellis Avenue, Chicago, IL 60637 \\
\EFIChicago Enrico Fermi Institute, University of Chicago, 5640 South Ellis Avenue, Chicago, IL 60637 \\
\PhysicsUChicago Department of Physics, University of Chicago, 5640 South Ellis Avenue, Chicago, IL 60637 \\
\ANL Argonne National Laboratory, 9700 S. Cass Avenue, Argonne, IL, USA 60439 \\
\Miss Department of Physics and Astronomy, University of Missouri, 5110 Rockhill Road, Kansas City, MO 64110 \\
\AAUChicago Department of Astronomy and Astrophysics, University of Chicago, 5640 South Ellis Avenue, Chicago, IL 60637 \\
\NIST NIST Quantum Devices Group, 325 Broadway Mailcode 817.03, Boulder, CO, USA 80305 \\
\PUC Departamento de Astronomia y Astrosifica, Pontificia Universidad Catolica,Chile \\
\McGill Department of Physics,McGill University, 3600 Rue University, Montreal, Quebec H3A 2T8, Canada \\
\illast Astronomy Department, University of Illinois at Urbana-Champaign,1002 W.\ Green Street,Urbana, IL 61801 USA \\
\illphy Department of Physics, University of Illinois Urbana-Champaign,1110 W.\ Green Street,Urbana, IL 61801 USA \\
\Berkeley Department of Physics, University of California, Berkeley, CA 94720 \\
\UFlorida Department of Astronomy, University of Florida, Gainesville, FL 32611 \\
\Colorado Department of Astrophysical and Planetary Sciences and Department of Physics, University of Colorado,Boulder, CO 80309 \\
\LBNL Physics Division, Lawrence Berkeley National Laboratory, Berkeley, CA 94720 \\
\Caltech California Institute of Technology, 1200 E. California Blvd., Pasadena, CA 91125 \\
\Arizona Steward Observatory, University of Arizona, 933 North Cherry Avenue, Tucson, AZ 85721 \\
\MIT Kavli Institute for Astrophysics and Space Research, Massachusetts Institute of Technology, 77 Massachusetts Avenue,
Cambridge, MA 02139 \\
\Michigan Department of Physics, University of Michigan, 450 Church Street, Ann Arbor, MI, 48109 \\
\CaseWestern Physics Department, Center for Education and Research in Cosmology and Astrophysics, Case Western Reserve University, Cleveland, OH 44106 \\
\Minnesota Physics Department, University of Minnesota, 116 Church Street S.E., Minneapolis, MN 55455 \\
\STScI Space Telescope Science Institute, 3700 San Martin Dr., Baltimore, MD 21218 \\
\SAIC Liberal Arts Department, School of the Art Institute of Chicago, 112 S Michigan Ave, Chicago, IL 60603 \\
\Dunlap Dunlap Institute for Astronomy \& Astrophysics, University of Toronto, 50 St George St, Toronto, ON, M5S 3H4, Canada \\
\Toronto Department of Astronomy \& Astrophysics, University of Toronto, 50 St George St, Toronto, ON, M5S 3H4, Canada \\
}
\begin{document}
\date{Accepted ???. Received ???; in original form ???}   

\maketitle                                                 
                                                           
\label{firstpage}      

\begin{abstract} 
The adiabatic evolution of the temperature of the cosmic microwave
background (CMB) is a key prediction of standard cosmology. We study
deviations from the expected adiabatic evolution of the CMB
temperature of the form $T(z) =T_0(1+z)^{1-\alpha}$ using measurements
of the spectrum of the Sunyaev Zel'dovich Effect with the South Pole
Telescope (SPT). We present a method for using the ratio of the
Sunyaev Zel'dovich signal measured at 95 and 150 GHz in the SPT data
to constrain the temperature of the CMB. We demonstrate that this
approach provides unbiased results using mock observations of clusters
from a new set of hydrodynamical simulations. We apply this method to
a sample of 158 SPT-selected clusters, spanning the redshift range
$0.05 < z < 1.35$, and measure $\alpha = 0.017^{+0.030}_{-0.028}$,
consistent with the standard model prediction of $\alpha=0$. In
combination with other published results, we constrain $\alpha = 0.011
\pm 0.016$, an improvement of $\sim 20\%$ over published
constraints. This measurement also provides a strong constraint on the
effective equation of state in models of decaying dark energy
$w_\mathrm{eff} = -0.987^{+0.016}_{-0.017}$.
\end{abstract}

\section{Introduction} 
\label{sec:intro} 

The existence of the cosmic microwave background is a fundamental
prediction of the Hot Big Bang Theory.  The intensity spectrum of the
CMB radiation locally has been measured by the COBE FIRAS instrument
and found to have a nearly exact blackbody spectrum with a temperature
of $T_0 = 2.72548 \pm 0.00057$ K (\citealt{fixsen09}).

A second fundamental prediction of the hot Big Bang theory is that the
CMB temperature must evolve over cosmic time.  Specifically, it is
expected to evolve as $T(z) =T_0(1+z)$, under the assumption that the
CMB photon fluid reacts adiabatically to the expansion of the Universe
as described by general relativity and electromagnetism. Deviations
from the adiabatic evolution of $T(z)$ would imply either a violation
of the hypothesis of local position invariance, and therefore of the
equivalence principle, or that the number of photons is not
conserved. In the former case, this could be associated with
variations of dimensionless coupling constants like the fine-structure
constant (see, e.g., \citealt{martins02}, \citealt{murphy03},
\citealt{srianand04}). The latter case is a consequence of many
physical processes predicted by non-standard cosmological models, such
as decaying vacuum energy density models, coupling between photons and
axion-like particles, and modified gravity scenarios.
\citep[e.g.,][]{mathiasek95,overduin98,lima00,puy04,jaeckel10,jetzer11a}.
In all of these models, energy has to be slowly injected or removed
from the CMB without distorting the Planck Spectrum sufficiently to
violate constraints from FIRAS \citep[]{avgoustidis12}.

Observational tests of non-standard temperature evolution typically
are parametrized by very simple models for the deviation. In
particular, we consider here the scaling law proposed by
\citet{lima00}: $ T(z) = T_0(1+z)^{1-\alpha}$, with $\alpha$ being a
free constant parameter\footnote{In previous literature this parameter
has been referred to with the Greek letter $\alpha$ or $\beta$. To
avoid confusion with the variable $\beta = v/c$ defined in
Eq. \ref{eq:sz}, we use $\alpha$.}. This is the phenomenological
parametrization that has been most widely studied by previous authors;
deviations of $\alpha$ from zero would result as a consequence of one
of the scenarios described above, such as the non-conservation of
photon number.

To date, two different observables have been used to determine $T(z)$.
At intermediate redshifts ($z \lesssim 1.5$), $T(z)$ can be determined
from measurements of the spectrum of the Sunyaev-Zel'dovich effect
(SZE) (\citealt{sunyaev72}), a technique first suggested by
\citet[]{fabbri78,rephaeli80}. The first attempt to measure $T(z)$
using the spectrum of the SZE was reported in \citet{battistelli02}
using multi-frequency observations of the clusters A2163 and
Coma. \citet{luzzi09} reported results from the analysis of a sample
of 13 clusters with $0.23\le z \le 0.546$. Adopting a flat prior on
$\alpha \in [0,1]$, they provided constraints $\alpha =
0.024^{+0.068}_{-0.024}$, consistent with standard adiabatic
evolution.

At high redshift ($z \gtrsim 1$), the CMB temperature can be
determined from quasar absorption line spectra which show atomic or
molecular fine structure levels excited by the photo-absorption of the
CMB radiation. If the system is in thermal equilibrium with the CMB,
then the excitation temperature of the energy states gives the
temperature of the black-body radiation
\citep[e.g.,][]{srianand00,molaro02,srianand08}. For example,
\citet{noterdaeme11} have reported on a sample of five carbon monoxide
absorption systems up to $z\sim 3$ where the CMB temperature has been
measured. They used their sample, in combination with low redshift SZE
measurements to place constraints on the phenomenological parameter
$\alpha = -0.007 \pm 0.027$. This also allowed them to put strong
constraints on the effective equation of state of decaying dark energy
models $w_\mathrm{eff} = -0.996 \pm 0.025$.  Recently,
\citet{avgoustidis12} extended this analysis by including constraints
inferred from differences between the angular diameter and luminosity
distances (the so-called distance-duality relation), which is also
affected in models in which photons can be created or destroyed. They
also showed that by releasing the positive prior assumption on
$\alpha$ the same cluster sample studied in \citet{luzzi09} constrains
$\alpha = 0.065 \pm 0.080$.

More recently, \citet{muller13} fit molecular absorption lines towards
quasars to measure the CMB temperature with an accuracy of a few
percent at $z=0.89$. Combining their data with the data
presented in \citet{noterdaeme11} they were able to further constrain
$\alpha = 0.009 \pm 0.019$.

Constraints on the CMB redshift evolution can be significantly
improved by including measurements of the SZE spectrum from
experiments, such as the South Pole Telescope (SPT) and Planck, with
much larger cluster samples. For instance, \citet{demartino12}
forecast the constraining power of Planck (further discussed in
\ref{sec:conclusions}) to measure $\alpha$. Using only clusters at $z
< 0.3$, they predicted that Planck could measure $\alpha$ with an
accuracy $\sigma_\alpha = 0.011$.

In this work, we present constraints on the temperature evolution of
the CMB using SZE spectral measurements at the 95 and 150 GHz bands
from the South Pole Telescope. The SPT is a 10m millimetre-wave
telescope operating at the South Pole \citep{carlstrom11} that has
recently completed a 2500 deg$^2$ multi-frequency survey of the
southern extragalactic sky.  Here we focus on the SZE selected cluster
sample that lies within a 720 deg$^2$ subregion where optical
follow-up and redshift measurements are complete
\citep{song12,reichardt12}.

\section{Method}
\label{sec:method}

Inverse Compton scattering of the CMB photons by the hot intracluster
medium induces secondary CMB temperature anisotropies in the direction
of clusters of galaxies. Neglecting relativistic corrections, the
thermal (tSZE) and kinematic (kSZE) contribution to the temperature
anisotropy in the direction $\hat{n}$ of a cluster at a frequency
$\nu$ can be approximated by (\citealt{demartino12}): \be \Delta
T(\hat{n}, \nu) \simeq T_0(\hat n)[ G(\nu)y_{\rm{c}}(\hat{n}) - \tau \beta].
\label{eq:sz}
\ee Here $T_0 (\hat n)$ is the current CMB temperature at the
direction $\hat n$, $\beta$ is the line of sight velocity of the
cluster in the CMB frame in units of the speed of light $c$ and $\tau$
is the optical depth. The Comptonization parameter $y_c$ is related to
the integrated pressure along the line of sight
$y_{\rm{c}}=(k_{\rm{B}}\sigma_{\rm{T}}/m_{\rm{e}}c^2)\int n_{\rm{e}}T_{\rm{e}} dl$ (where $n_{\rm{e}}$ and $T_{\rm{e}}$ are
respectively the electron density and temperature). In the
non-relativistic regime and for adiabatic expansion, $G(x)= x\,{\rm
coth}(x/2)-4$, where the reduced frequency $x$ is given by
$x=h\nu(z)/k_{\rm{B}}T(z)=h\nu_0(1+z) / [k_{\rm{B}}T_0(1+z)] \equiv x_0$ and is
independent of redshift, $\nu(z)$ is the frequency of a CMB photon
scattered by the intra-cluster medium and $T(z)$ is the black body
temperature of the CMB at the cluster location.

If $T(z) = T_0(1+z)^{1-\alpha}$, then the reduced frequency varies as
$x(z,\alpha)=x_0(1+z)^\alpha$ and the spectral frequency dependence of
$G(\nu)$, the tSZE, now also depends on $\alpha$:
$G(x)=G(\nu_0,\alpha, z)$.  From Eq. \ref{eq:sz}, neglecting the kSZE
contribution, it follows that measuring the ratio of temperature decrements at
two different frequencies $\nu_1$ and $\nu_2$ provides: \be
R(\nu_1,\nu_2,z,\alpha) \equiv \frac{\Delta T(\hat{n},\nu_1,z)}{\Delta
T(\hat{n},\nu_2,z) } \simeq
\frac{G(\nu_1,z,\alpha)}{G(\nu_2,z,\alpha)} \label{eq:ratio} \ee This
ratio is redshift independent for $\alpha = 0$, but not in the case of
$\alpha \neq 0$. This method has the advantage that, by taking ratios,
the dependence on the Comptonization parameter $y_c$ (and therefore on
the cluster properties) is removed and the need to account for model
uncertainties on the gas density and temperature profile is avoided
(Battistelli et al 2002, Luzzi et al 2009). Note that in this approach
the distribution of temperature ratios is, in general, non-Gaussian
(Luzzi et al. 2009) and needs to be properly modelled.

One important source of noise in these measurements is the primary
anisotropy of the CMB. To precisely measure $\Delta T(\hat{n},\nu)$
for a single cluster, we would have to remove the primary CMB
anisotropies in the direction $\hat{n}$. In principle, this could be
done by subtracting the CMB temperature measured near the SZE null
frequency, which, in the case of $\alpha = 0$ and non-relativistic
ICM, is given by a map obtained at 217 \gz (\citealt{demartino12}).
Alternatively, because the primary CMB fluctuations are random, it is
possible reduce this source of noise by averaging over a large sample
of clusters.

In \citet{reichardt12}, the SPT cluster sample was selected using a
matched multi-frequency spatial filter \citep{melin06}, designed to
optimally measure the cluster signal given knowledge of the cluster
profile and the noise in the maps. The cluster gas
profiles are assumed to be well fit by a spherical $\beta$ model
\citep{cavaliere76}, with $\beta=1$ and twelve possible core radii,
$\theta_c$, linearly spaced from 0.25 to 3 arcmin. The noise
contributions include astrophysical (e.g., the CMB, point sources) and
instrumental (e.g., atmospheric, detector) contributions.  For each
cluster, the maximum signal to noise in the spatially filtered maps
was denoted as $\xi$.

In this work, we measure the ratio of the CMB temperature decrements
in the SPT data at 95 and 150 GHz.  We extract the cluster signal from
the single-frequency spatially filtered maps at 95 and 150 GHz, using
the SPT position and core radius favored by the multi-frequency
analysis in \citet{reichardt12}.  To compare the decrement at each
frequency, we need to account for the smaller beam at 150 GHz.  We do
this by convolving the 150~GHz data to the same beam size as the 95
GHz data, and then using the 95~GHz filter to extract the signal from
the resultant 150~GHz maps.  The associated uncertainty in the CMB
temperature decrement would be equal to the R.M.S. of the
single-frequency spatially filtered maps.

Finally, we use the derived values of temperature in the two bands and
the associated cluster redshift to constrain $\alpha$ from
Eq. \ref{eq:ratio} through a maximum likelihood analysis (Luzzi et
al. 2009) where the likelihood is defined as : \be {\cal L}(\alpha)
\propto \prod \limits_{i=1}^{N_\mathrm{clus}} \exp \left\{-\frac{
(T^{(i)}_{150} R(z^{(i)},\alpha) - T^{(i)}_{95})^2}{2[(\Delta
T^{(i)}_{150} R(z^{(i)},\alpha))^2 + (\Delta T^{(i)}_{95})^2]}
\right\}
\label{eq:likely},
\ee and $R(z,\alpha) \equiv R(95$~GHz$,150$~GHz$,z,\alpha)$ according
to Eq. \ref{eq:ratio} is calculated by integrating:\be R(z,\alpha) =
\frac{\int{G(\nu,z,\alpha)F_{95}(\nu)
d\nu}}{\int{G(\nu,z,\alpha)F_{150}(\nu) d\nu}},\ee where $F_{95}$ and
$F_{150}$ are the measured filter response of the SPT 95 and 150 GHz
bands, normalized such that the integral over each of the bands is
one. We have assumed the non-relatitivistic expression for
$G(\nu,z,\alpha)$, however, we find relativistic corrections have a
negligible effect on our result.  For the range of electron
temperatures expected in our cluster sample, including relativistic
corrections from \citet{itoh98} changes our final constraints on
$\alpha$ by less than 1 per cent.

\section{Verification of Method with Simulations}
\label{sec:sim}

We test the method described above using simulations. To do so, we
make mock SPT observations of clusters that are formed in a large
volume, high resolution cosmological hydrodynamical simulation (Dolag
et al., in preparation). The simulation has been carried out with
P-GADGET3, a modification of P-GADGET-2 (Springel 2005). The code uses
an entropy-conserving formulation of SPH (Springel \& Hernquist 2002)
and includes treatment of radiative cooling, heating by a UV
background, star formation and feedback processes from supernovae
explosions and active galactic nuclei \citep[]{springel03,
fabjan10}. Cosmological parameters are chosen to match WMAP7
(\citealt{komatsu11}). The simulation box is 1244~Mpc per side and
contains $1526^3$ dark matter particles and as many gas particles,
from which five simulated SZE light-cones, each of size $13^\circ
\times 13^\circ $ (i.e., the total solid angle is 845 deg$^2$) have
been extracted up to $z \sim 2$. From each of these simulated SZE
maps, we then construct simulated SPT maps at 95 and 150 GHz.

The details of the construction of the simulated SZE light-cones will
be presented in a forthcoming paper (Liu et al., in preparation), we
highlight here the basic properties. In these mock observations, we
include contributions from: (1) primary CMB anisotropies, (2)
convolution with the SPT 150~GHz and 95~GHz beams, (3) instrumental
noise consistent with the observed SPT map depths of 18 and 44
$\mu$K-arcmin for the 150 GHz and 95 GHz bands, respectively, and (4)
associated filter transfer functions for the two simulated
bands. Finally, from these mock maps we identify clusters with the
same approach adopted for real SPT clusters
\citep[e.g.,][]{staniszewski09,reichardt12}, obtaining a sample of 212
clusters above signal to noise $\xi = 4.5$.

We then measure the ratio of the temperatures in the two bands, using
the approach described in Section 2. We first convolve the 150 GHz
maps to match the larger beam of the 95 GHz band. We then individually
filter the 95~GHz and the 150~GHz maps with the 95~GHz filter and
measure the signal at the position and $\theta_C$ scale that maximize
the signal to noise in the multifrequency analysis. We then maximize
the likelihood to determine $\alpha$ (Eq. \ref{eq:likely}). We recover
$\alpha = 0.0019 \pm 0.022$, in agreement with the input value of
$\alpha = 0$.

\section{SPT Results}
\label{sec:SPT}

We measure the temperature decrement ratios at the positions of the
SPT-selected cluster sample from \citep{reichardt12}, which included
data from 720 of the 2500 deg$^2$ SPT-SZ survey. The SPT-SZ data used
here has typical noise levels of 44~$\mu$K-arcmin and 18~$\mu$K-arcmin
in CMB temperature units at 95~GHz and 150~GHz, respectively.  The
exceptions are the two fields centered at $23^{h}30^{m},-55^{d}$ and
$5^h30^m,-55^d$ from \citet{reichardt12}, which have a depth of
13~$\mu$K-arcmin in the 150~GHz data used in this work.  Since the
publication of \citet{reichardt12}, these fields had been re-observed
in the SPT-SZ survey, with the new observations providing new 95 GHz
measurements and deeper 150~GHz data.  The final cluster sample used
here consists of 158 clusters with both a $\xi>4.5$ from
\citet{reichardt12}, and either a spectroscopic or photometric
redshift reported in \citet{song12}.  We refer the reader to
\citet{staniszewski09}, \citet{vanderlinde10}, \citet{schaffer11},
\citet{williamson11}, \citet{reichardt12} for a detailed description of
the survey strategy and dataset characteristics .

We apply the same technique described in Section \ref{sec:method} and
tested in Section \ref{sec:sim} to measure the evolution of the CMB
temperature with SPT clusters. 

Using SPT data alone, we constrain the temperature evolution of the CMB to be  \be \alpha =
0.017^{+0.030}_{-0.028},
\label{eq:spt_results}
\ee which is consistent with the adiabatic expectation of $\alpha=0$.
We estimate the instrumental uncertainties associated with the beams,
calibration, and filter responses, and find them all to have a
negligible result on this constraint. Moreover, the statistical
uncertainty is $\sim 30 \%$ larger than the limit on possible
observational biases implied by the results of Section \ref{sec:sim},
implying that our analysis method is shown to be unbiased at or below
the level of the statistical uncertainty.

\begin{figure}
 \hbox{\psfig{file=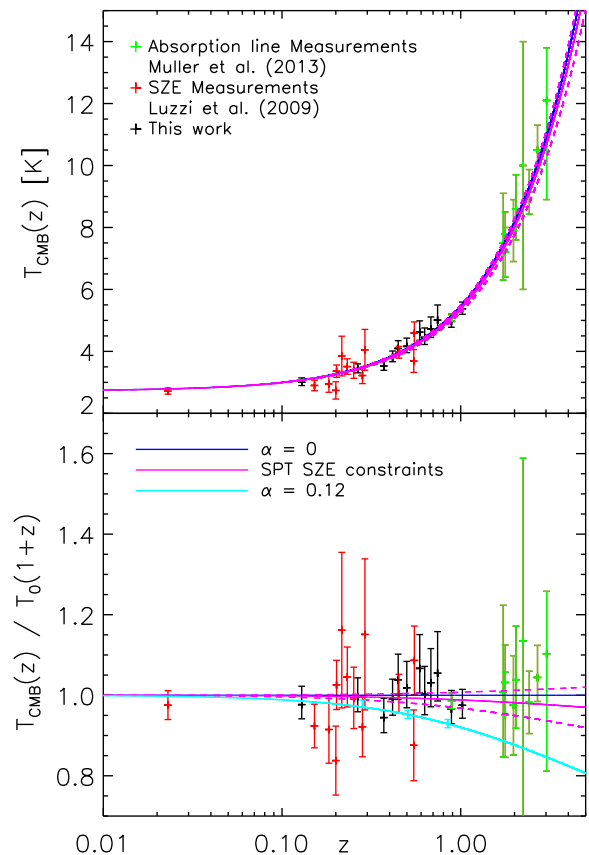,width=9.0cm}}
 \caption{{\it Top panel:} Measurements of the temperature of the CMB
   as a function of redshift. Red points correspond to SZE
   measurements toward galaxy clusters (see Luzzi et al. 2009 and
   references therein), green points are absorption lines studies (see
   Muller et al 2013 and references therein). Black points are the SPT
   SZE cluster constraints. For visualization purposes SPT clusters
   results have been stacked in 12 equally populated redshift
   bins. The blue continuous line corresponds to the relation $T(z) =
   T_0 \times (1+ z)$ and solid and dashed purple lines are the
   evolution corresponding to the best fit and $\pm$1$\sigma$
   models. {\it Bottom panel:} Deviation of the measured temperature
   of the CMB as a function of redshift with respect to the adiabatic
   evolution. Cyan points represent the measured temperature of the
   CMB in three stacked redshift bins for a simulation with input
   value $\alpha = 0.12$ (cyan solid line).}
 \label{fi:t_z}
\end{figure}

We further combine our results with previously published data (Fig
\ref{fi:t_z}). In particular we include measurements from clusters
collected by Luzzi et al. (2009) and fine structure absorption line
studies collected by Noterdaeme et al. (2011) and Muller et
al. (2013).  We thus obtain a tighter constraint on the $T(z) = T_0 (1
+ z)^{(1-\alpha)}$ law: \be \alpha = 0.011 \pm 0.016,
\label{eq:spt_ext}
\ee a $ \sim 20\%$ improvement in measurement uncertainty in
comparison to the previously reported $\alpha = 0.009 \pm 0.019$
(Muller et al. 2013). We also note that previous results based on SZE
measurments \citep[]{luzzi09} have a negligible impact in this
constraint.

The measurement presented here is consistent with the adiabatic
evolution of the CMB radiation temperature ($\alpha=0$) expected from
the standard hot Big-Bang model. Considering alternative cosmological
models, \citet{jetzer11} demonstrated that measuring $T(z)$ at
different redshifts allows one to constrain the effective equation of
state of decaying dark energy ($p=w_\mathrm{eff} \rho$). Following
\citet{noterdaeme11}, by fitting the combined constraints on $T(z)$
with the temperature-redshift relation (Eq. 22 in \citealt{jetzer11}),
taking $\Omega_{\rm{m}}=0.255 \pm 0.016$ \citep{reichardt12} and
fixing the adiabatic index $\gamma$ to the canonical value (4/3), we
get $w_\mathrm{eff}=-0.987^{+0.016}_{-0.017}$, in comparison with
$w_\mathrm{eff}=-0.996 \pm 0.025$ obtained by \citet{noterdaeme11}.

\subsection{Selection bias}
\label{sec:selection bias}
A number of possible selection biases could affect our
measurements. In particular, cluster candidates were identified using
a multi-band matched-filter approach (\citealt{melin06}) where the
temperature evolution of the Universe is assumed to be adiabatic. This
could therefore bias our selection towards clusters that best mimic
this behavior. To show that this is not the case, we construct SPT
mock lightcones similar to the ones presented in Sect. \ref{sec:sim}
but assuming different values of $\alpha$. We then performed the same
analysis described in Sect. \ref{sec:sim} and show that we are able to
recover the input value. Specifically we test simulations with input
values of $\alpha$ offset by more than 3$\sigma$ from the adiabatic
value, $\alpha = -0.12$ and $\alpha = 0.12$. We then select clusters
with the above described matched-filter multi-frequency cluster finder
under the assumption of adiabatic evolution and constrain $\alpha$. We
obtain unbiased measurements for the underlying input value $\alpha =
-0.111^{+0.022}_{-0.018}$ and $\alpha = 0.110^{+0.014}_{-0.014}$, thus
demonstrating that the selection is not driving our constraints
(bottom panel of Fig. \ref{fi:t_z}).

Another potential source of bias in our measurement of $\alpha$ is the
fact that the temperature fluctuations of the CMB at the location of
the SPT clusters should not average to zero. In fact, due to the
adopted cluster selection, negative temperature fluctuations are more
likely than positive ones \citep[]{vanderlinde10}. We estimate this
effect to be negligible using the simulations described in
Sect. \ref{sec:sim}. We also note that this effect should be less
significant at larger SPT signal to noise $\xi$ \citep[]{benson13}.
If we restrict our analysis to the clusters with $\xi > 8$, which
reduces the cluster sample by a factor of $\sim$6 to the 24 highest
signal-to-noise clusters, we constrain $\alpha =
0.023^{+0.044}_{-0.038}$.  This is consistent with our main result
with only a modest $30\%$ increase in the uncertainty in $\alpha$.
This demonstrates that the constraints depend most significantly on
the highest signal to noise clusters, which will be less biased by the
CMB from the SPT-selection. Similarly, we estimate the bias associated
with lensed dusty sources to be unimportant for our analysis; their
primary impact would be introducing some skewness in the scatter of
clusters about our best fit model \citep[]{hezaveh13}.

Emission from cluster galaxies can also potentially bias our
measurement. We estimate the effect to be negligible by performing the
analysis presented here on subsamples of clusters above different
$\xi$ thresholds and by excluding clusters in proximity to known SUMSS
sources \citep[]{mauch03}. All subsamples examined provide
statistically consistent results. For example, using a subsample of 75
clusters with no associated SUMSS sources brighter than 20 mJy within
a projected distance of 3 arcmin from the cluster centers, we obtain
consistent results of $\alpha = 0.021^{+0.042}_{-0.038}$.

\section{Conclusions}
\label{sec:conclusions}
We have studied deviations from the adiabatic evolution of the mean
temperature of the CMB in the form of $T(z) = T_0(1+z)^{(1-\alpha)}$.
We present a method based on matched-filtering of clusters at the SPT
frequencies and show that we are able to recover unbiased results
using simulated clusters. The simulated lightcones we use come from a
large cosmological hydrodynamical simulation and include realistic SPT
beam effects, CMB anisotropy and SPT noise levels for both the 150~\gz
and 95~\gz bands.

We apply this method to a sample of 158 SPT clusters selected from 720
square degrees of the 2500 square degree SPT-SZ survey, which
span the redshift range $0.05<z<1.35$, and measure $\alpha =
0.017^{+0.030}_{-0.028}$, consistent with the standard model
prediction of $\alpha=0$. Our measurement gives competitive
constraints and significantly extends the redshift range with respect
to previously published results based on galaxy clusters
\citep[e.g.,][]{luzzi09,avgoustidis12,demartino12,muller13}.  Combining
our measurements with published data we obtain $\alpha = 0.011 \pm
0.016$, a $20\%$ improvement with respect to current published
constraints.

Such tight limits on deviations from the adiabatic evolution of the
CMB also put interesting constraints on the effective equation of
state of decaying dark energy models, $w_\mathrm{eff}$. Indeed, from
SPT clusters alone we are able to measure
$w_\mathrm{eff}=-0.988^{+0.029}_{-0.033}$, in good agreement with
previous constraints based on quasar absorption lines
\citep[]{noterdaeme11}.

Coincident with the submission of this analysis, \citet{Hurier13}
released results of a similar analysis carried out on 1839 galaxy
clusters observed with Planck. The cluster sample they adopted also
included the SPT sample that we analyse here, although it did not
contribute significantly to their main results. They were able to
constrain $\alpha = 0.009 \pm 0.017$ by stacking the Planck catalog of
SZE detected clusters \citep{planck_xxix} in different redshift bins,
with only one cluster in each of their highest redshift bins $z = 0.8$
and $z = 1$. Because the SPT data are on average a factor of 3 deeper
than Planck, and the SPT beam is $\sim8$ times smaller, the SPT
dataset provides stronger constraints on a per cluster basis and is
particularly well suited for studies of the high redshift tail of the
cluster distribution.

Future analyses will be able to draw upon larger cluster samples
(e.g., the full 2500 square degree SPT-SZ survey and the upcoming
SPTpol and SPT-3G surveys) and quasar surveys (e.g., SDSS III). By
expanding the data volume at high redshifts, these surveys will enable
precision tests of the temperature evolution of the CMB across cosmic
time.  Moreover, because clusters and quasars suffer from different
systematics, the comparison will provide an important cross-check on
systematics. These surveys will improve constraints on non-standard
cosmological models.

The Munich SPT group is supported by the DFG through TR33 ``The Dark
Universe'' and the Cluster of Excellence ``Origin and Structure of the
Universe''.  The South Pole Telescope program is supported by the
National Science Foundation through grant ANT-0638937. Partial support
is also provided by the NSF Physics Frontier Center grant PHY-0114422
to the Kavli Institute of Cosmological Physics at the University of
Chicago, by the Kavli Foundation and the Gordon and Betty Moore
Foundation. Galaxy cluster research at Harvard is supported by NSF
grants AST-1009012 and DGE-1144152. Galaxy cluster research at SAO is
supported in part by NSF grants AST-1009649 and MRI-0723073. The
McGill group acknowledges funding from the National Sciences and
Engineering Research Council of Canada, Canada Research Chairs
program, and the Canadian Institute for Advanced Research.
\bibliographystyle{mn2e}
\bibliography{paper,spt}

\begin{thebibliography}{41}
\expandafter\ifx\csname natexlab\endcsname\relax\def\natexlab#1{#1}\fi

\bibitem[{{Avgoustidis} {et~al}\mbox{.}(2012){Avgoustidis}, {Luzzi}, {Martins},
  \& {Monteiro}}]{avgoustidis12}
{Avgoustidis} A., {Luzzi} G., {Martins} C.~J.~A.~P., {Monteiro} A.~M.~R.~V.~L.,
  2012, \jcp, 2, 13

\bibitem[{{Battistelli} {et~al}\mbox{.}(2002){Battistelli}, {De Petris},
  {Lamagna}, {Melchiorri}, {Palladino}, {Savini}, {Cooray}, {Melchiorri},
  {Rephaeli}, \& {Shimon}}]{battistelli02}
{Battistelli} E.~S. {et~al.}, 2002, \apjl, 580, L101

\bibitem[{{Benson} {et~al}\mbox{.}(2013){Benson}, {de Haan}, {Dudley},
  {Reichardt}, {Aird}, {Andersson}, {Armstrong}, {Ashby}, {Bautz}, {Bayliss},
  {Bazin}, {Bleem}, {Brodwin}, {Carlstrom}, {Chang}, {Cho}, {Clocchiatti},
  {Crawford}, {Crites}, {Desai}, {Dobbs}, {Foley}, {Forman}, {George},
  {Gladders}, {Gonzalez}, {Halverson}, {Harrington}, {High}, {Holder},
  {Holzapfel}, {Hoover}, {Hrubes}, {Jones}, {Joy}, {Keisler}, {Knox}, {Lee},
  {Leitch}, {Liu}, {Lueker}, {Luong-Van}, {Mantz}, {Marrone}, {McDonald},
  {McMahon}, {Mehl}, {Meyer}, {Mocanu}, {Mohr}, {Montroy}, {Murray}, {Natoli},
  {Padin}, {Plagge}, {Pryke}, {Rest}, {Ruel}, {Ruhl}, {Saliwanchik}, {Saro},
  {Sayre}, {Schaffer}, {Shaw}, {Shirokoff}, {Song}, {Spieler}, {Stalder},
  {Staniszewski}, {Stark}, {Story}, {Stubbs}, {Suhada}, {van Engelen},
  {Vanderlinde}, {Vieira}, {Vikhlinin}, {Williamson}, {Zahn}, \&
  {Zenteno}}]{benson13}
{Benson} B.~A. {et~al.}, 2013, \apj, 763, 147

\bibitem[{{Carlstrom} {et~al}\mbox{.}(2011){Carlstrom}, {Ade}, {Aird},
  {Benson}, {Bleem}, {Busetti}, {Chang}, {Chauvin}, {Cho}, {Crawford},
  {Crites}, {Dobbs}, {Halverson}, {Heimsath}, {Holzapfel}, {Hrubes}, {Joy},
  {Keisler}, {Lanting}, {Lee}, {Leitch}, {Leong}, {Lu}, {Lueker}, {Luong-van},
  {McMahon}, {Mehl}, {Meyer}, {Mohr}, {Montroy}, {Padin}, {Plagge}, {Pryke},
  {Ruhl}, {Schaffer}, {Schwan}, {Shirokoff}, {Spieler}, {Staniszewski},
  {Stark}, {Tucker}, {Vanderlinde}, {Vieira}, \& {Williamson}}]{carlstrom11}
{Carlstrom} J.~E. {et~al.}, 2011, \pasp, 123, 568

\bibitem[{{Cavaliere} \& {Fusco-Femiano}(1976)}]{cavaliere76}
{Cavaliere} A., {Fusco-Femiano} R., 1976, \aap, 49, 137

\bibitem[{{de Martino} {et~al}\mbox{.}(2012){de Martino}, {Atrio-Barandela},
  {da Silva}, {Ebeling}, {Kashlinsky}, {Kocevski}, \& {Martins}}]{demartino12}
{de Martino} I., {Atrio-Barandela} F., {da Silva} A., {Ebeling} H.,
  {Kashlinsky} A., {Kocevski} D., {Martins} C.~J.~A.~P., 2012, \apj, 757, 144

\bibitem[{{Fabbri} {et~al}\mbox{.}(1978){Fabbri}, {Melchiorri}, \&
  {Natale}}]{fabbri78}
{Fabbri} R., {Melchiorri} F., {Natale} V., 1978, \apss, 59, 223

\bibitem[{{Fabjan} {et~al}\mbox{.}(2010){Fabjan}, {Borgani}, {Tornatore},
  {Saro}, {Murante}, \& {Dolag}}]{fabjan10}
{Fabjan} D., {Borgani} S., {Tornatore} L., {Saro} A., {Murante} G., {Dolag} K.,
  2010, \mnras, 401, 1670

\bibitem[{{Fixsen} {et~al}\mbox{.}(2009){Fixsen}, {Kogut}, {Levin}, {Limon},
  {Lubin}, {Mirel}, {Seiffert}, {Singal}, {Wollack}, {Villela}, \&
  {Wuensche}}]{fixsen09}
{Fixsen} D.~J. {et~al.}, 2009, submitted to \apj, astro-ph/0901.0555

\bibitem[{{Hezaveh} {et~al}\mbox{.}(2013){Hezaveh}, {Vanderlinde}, {Holder}, \&
  {de Haan}}]{hezaveh13}
{Hezaveh} Y., {Vanderlinde} K., {Holder} G., {de Haan} T., 2013, \apj, 772, 121

\bibitem[{{Hurier} {et~al}\mbox{.}(2013){Hurier}, {Aghanim}, {Douspis}, \&
  {Pointecouteau}}]{Hurier13}
{Hurier} G., {Aghanim} N., {Douspis} M., {Pointecouteau} E., 2013, ArXiv
  e-prints:1311.4694

\bibitem[{{Itoh} {et~al}\mbox{.}(1998){Itoh}, {Kohyama}, \& {Nozawa}}]{itoh98}
{Itoh} N., {Kohyama} Y., {Nozawa} S., 1998, \apj, 502, 7

\bibitem[{{Jaeckel} \& {Ringwald}(2010)}]{jaeckel10}
{Jaeckel} J., {Ringwald} A., 2010, Annual Review of Nuclear and Particle
  Science, 60, 405

\bibitem[{{Jetzer} {et~al}\mbox{.}(2011){Jetzer}, {Puy}, {Signore}, \&
  {Tortora}}]{jetzer11}
{Jetzer} P., {Puy} D., {Signore} M., {Tortora} C., 2011, General Relativity and
  Gravitation, 43, 1083

\bibitem[{{Jetzer} \& {Tortora}(2011)}]{jetzer11a}
{Jetzer} P., {Tortora} C., 2011, \prd, 84, 043517

\bibitem[{{Komatsu} {et~al}\mbox{.}(2011){Komatsu}, {Smith}, {Dunkley},
  {Bennett}, {Gold}, {Hinshaw}, {Jarosik}, {Larson}, {Nolta}, {Page},
  {Spergel}, {Halpern}, {Hill}, {Kogut}, {Limon}, {Meyer}, {Odegard}, {Tucker},
  {Weiland}, {Wollack}, \& {Wright}}]{komatsu11}
{Komatsu} E. {et~al.}, 2011, \apjs, 192, 18

\bibitem[{{Lima} {et~al}\mbox{.}(2000){Lima}, {Silva}, \& {Viegas}}]{lima00}
{Lima} J.~A.~S., {Silva} A.~I., {Viegas} S.~M., 2000, \mnras, 312, 747

\bibitem[{{Luzzi} {et~al}\mbox{.}(2009){Luzzi}, {Shimon}, {Lamagna},
  {Rephaeli}, {De Petris}, {Conte}, {De Gregori}, \& {Battistelli}}]{luzzi09}
{Luzzi} G., {Shimon} M., {Lamagna} L., {Rephaeli} Y., {De Petris} M., {Conte}
  A., {De Gregori} S., {Battistelli} E.~S., 2009, \apj, 705, 1122

\bibitem[{{Martins}(2002)}]{martins02}
{Martins} C.~J.~A.~P., 2002, Royal Society of London Philosophical Transactions
  Series A, 360, 2681

\bibitem[{{Matyjasek}(1995)}]{mathiasek95}
{Matyjasek} J., 1995, \prd, 51, 4154

\bibitem[{{Mauch} {et~al}\mbox{.}(2003){Mauch}, {Murphy}, {Buttery}, {Curran},
  {Hunstead}, {Piestrzynski}, {Robertson}, \& {Sadler}}]{mauch03}
{Mauch} T., {Murphy} T., {Buttery} H.~J., {Curran} J., {Hunstead} R.~W.,
  {Piestrzynski} B., {Robertson} J.~G., {Sadler} E.~M., 2003, \mnras, 342, 1117

\bibitem[{{Melin} {et~al}\mbox{.}(2006){Melin}, {Bartlett}, \&
  {Delabrouille}}]{melin06}
{Melin} J.-B., {Bartlett} J.~G., {Delabrouille} J., 2006, \aap, 459, 341

\bibitem[{{Molaro} {et~al}\mbox{.}(2002){Molaro}, {Levshakov},
  {Dessauges-Zavadsky}, \& {D'Odorico}}]{molaro02}
{Molaro} P., {Levshakov} S.~A., {Dessauges-Zavadsky} M., {D'Odorico} S., 2002,
  \aap, 381, L64

\bibitem[{{Muller} {et~al}\mbox{.}(2013){Muller}, {Beelen}, {Black}, {Curran},
  {Horellou}, {Aalto}, {Combes}, {Gu{\'e}lin}, \& {Henkel}}]{muller13}
{Muller} S. {et~al.}, 2013, \aap, 551, A109

\bibitem[{{Murphy} {et~al}\mbox{.}(2003){Murphy}, {Webb}, \&
  {Flambaum}}]{murphy03}
{Murphy} M.~T., {Webb} J.~K., {Flambaum} V.~V., 2003, \mnras, 345, 609

\bibitem[{{Noterdaeme} {et~al}\mbox{.}(2011){Noterdaeme}, {Petitjean},
  {Srianand}, {Ledoux}, \& {L{\'o}pez}}]{noterdaeme11}
{Noterdaeme} P., {Petitjean} P., {Srianand} R., {Ledoux} C., {L{\'o}pez} S.,
  2011, \aap, 526, L7

\bibitem[{{Overduin} \& {Cooperstock}(1998)}]{overduin98}
{Overduin} J.~M., {Cooperstock} F.~I., 1998, \prd, 58, 043506

\bibitem[{{Planck Collaboration}(2013)}]{planck_xxix}
{Planck Collaboration}, 2013, ArXiv e-prints:1303.5089

\bibitem[{{Puy}(2004)}]{puy04}
{Puy} D., 2004, \aap, 422, 1

\bibitem[{{Reichardt} {et~al}\mbox{.}(2013){Reichardt}, {Stalder}, {Bleem},
  {Montroy}, {Aird}, {Andersson}, {Armstrong}, {Ashby}, {Bautz}, {Bayliss},
  {Bazin}, {Benson}, {Brodwin}, {Carlstrom}, {Chang}, {Cho}, {Clocchiatti},
  {Crawford}, {Crites}, {de Haan}, {Desai}, {Dobbs}, {Dudley}, {Foley},
  {Forman}, {George}, {Gladders}, {Gonzalez}, {Halverson}, {Harrington},
  {High}, {Holder}, {Holzapfel}, {Hoover}, {Hrubes}, {Jones}, {Joy}, {Keisler},
  {Knox}, {Lee}, {Leitch}, {Liu}, {Lueker}, {Luong-Van}, {Mantz}, {Marrone},
  {McDonald}, {McMahon}, {Mehl}, {Meyer}, {Mocanu}, {Mohr}, {Murray}, {Natoli},
  {Padin}, {Plagge}, {Pryke}, {Rest}, {Ruel}, {Ruhl}, {Saliwanchik}, {Saro},
  {Sayre}, {Schaffer}, {Shaw}, {Shirokoff}, {Song}, {Spieler}, {Staniszewski},
  {Stark}, {Story}, {Stubbs}, {{\v S}uhada}, {van Engelen}, {Vanderlinde},
  {Vieira}, {Vikhlinin}, {Williamson}, {Zahn}, \& {Zenteno}}]{reichardt12}
{Reichardt} C.~L. {et~al.}, 2013, \apj, 763, 127

\bibitem[{{Rephaeli}(1980)}]{rephaeli80}
{Rephaeli} Y., 1980, \apj, 241, 858

\bibitem[{{Schaffer} {et~al}\mbox{.}(2011){Schaffer}, {Crawford}, {Aird},
  {Benson}, {Bleem}, {Carlstrom}, {Chang}, {Cho}, {Crites}, {de Haan}, {Dobbs},
  {George}, {Halverson}, {Holder}, {Holzapfel}, {Hoover}, {Hrubes}, {Joy},
  {Keisler}, {Knox}, {Lee}, {Leitch}, {Lueker}, {Luong-Van}, {McMahon}, {Mehl},
  {Meyer}, {Mohr}, {Montroy}, {Padin}, {Plagge}, {Pryke}, {Reichardt}, {Ruhl},
  {Shirokoff}, {Spieler}, {Stalder}, {Staniszewski}, {Stark}, {Story},
  {Vanderlinde}, {Vieira}, \& {Williamson}}]{schaffer11}
{Schaffer} K.~K. {et~al.}, 2011, \apj, 743, 90

\bibitem[{{Song} {et~al}\mbox{.}(2012){Song}, {Zenteno}, {Stalder}, {Desai},
  {Bleem}, {Aird}, {Armstrong}, {Ashby}, {Bayliss}, {Bazin}, {Benson},
  {Bertin}, {Brodwin}, {Carlstrom}, {Chang}, {Cho}, {Clocchiatti}, {Crawford},
  {Crites}, {de Haan}, {Dobbs}, {Dudley}, {Foley}, {George}, {Gettings},
  {Gladders}, {Gonzalez}, {Halverson}, {Harrington}, {High}, {Holder},
  {Holzapfel}, {Hoover}, {Hrubes}, {Joy}, {Keisler}, {Knox}, {Lee}, {Leitch},
  {Liu}, {Lueker}, {Luong-Van}, {Marrone}, {McDonald}, {McMahon}, {Mehl},
  {Meyer}, {Mocanu}, {Mohr}, {Montroy}, {Natoli}, {Nurgaliev}, {Padin},
  {Plagge}, {Pryke}, {Reichardt}, {Rest}, {Ruel}, {Ruhl}, {Saliwanchik},
  {Saro}, {Sayre}, {Schaffer}, {Shaw}, {Shirokoff}, {{\v S}uhada}, {Spieler},
  {Stanford}, {Staniszewski}, {Stark}, {Story}, {Stubbs}, {van Engelen},
  {Vanderlinde}, {Vieira}, {Williamson}, \& {Zahn}}]{song12}
{Song} J. {et~al.}, 2012, \apj, 761, 22

\bibitem[{{Springel} \& {Hernquist}(2003)}]{springel03}
{Springel} V., {Hernquist} L., 2003, \mnras, 339, 289

\bibitem[{{Srianand} {et~al}\mbox{.}(2004){Srianand}, {Chand}, {Petitjean}, \&
  {Aracil}}]{srianand04}
{Srianand} R., {Chand} H., {Petitjean} P., {Aracil} B., 2004, Physical Review
  Letters, 92, 121302

\bibitem[{{Srianand} {et~al}\mbox{.}(2008){Srianand}, {Noterdaeme}, {Ledoux},
  \& {Petitjean}}]{srianand08}
{Srianand} R., {Noterdaeme} P., {Ledoux} C., {Petitjean} P., 2008, \aap, 482,
  L39

\bibitem[{{Srianand} {et~al}\mbox{.}(2000){Srianand}, {Petitjean}, \&
  {Ledoux}}]{srianand00}
{Srianand} R., {Petitjean} P., {Ledoux} C., 2000, \nat, 408, 931

\bibitem[{{Staniszewski} {et~al}\mbox{.}(2009){Staniszewski}, {Ade}, {Aird},
  {Benson}, {Bleem}, {Carlstrom}, {Chang}, {Cho}, {Crawford}, {Crites}, {de
  Haan}, {Dobbs}, {Halverson}, {Holder}, {Holzapfel}, {Hrubes}, {Joy},
  {Keisler}, {Lanting}, {Lee}, {Leitch}, {Loehr}, {Lueker}, {McMahon}, {Mehl},
  {Meyer}, {Mohr}, {Montroy}, {Ngeow}, {Padin}, {Plagge}, {Pryke}, {Reichardt},
  {Ruhl}, {Schaffer}, {Shaw}, {Shirokoff}, {Spieler}, {Stalder}, {Stark},
  {Vanderlinde}, {Vieira}, {Zahn}, \& {Zenteno}}]{staniszewski09}
{Staniszewski} Z. {et~al.}, 2009, \apj, 701, 32

\bibitem[{{Sunyaev} \& {Zel'dovich}(1972)}]{sunyaev72}
{Sunyaev} R.~A., {Zel'dovich} Y.~B., 1972, Comments on Astrophysics and Space
  Physics, 4, 173

\bibitem[{{Vanderlinde} {et~al}\mbox{.}(2010){Vanderlinde}, {Crawford}, {de
  Haan}, {Dudley}, {Shaw}, {Ade}, {Aird}, {Benson}, {Bleem}, {Brodwin},
  {Carlstrom}, {Chang}, {Crites}, {Desai}, {Dobbs}, {Foley}, {George},
  {Gladders}, {Hall}, {Halverson}, {High}, {Holder}, {Holzapfel}, {Hrubes},
  {Joy}, {Keisler}, {Knox}, {Lee}, {Leitch}, {Loehr}, {Lueker}, {Marrone},
  {McMahon}, {Mehl}, {Meyer}, {Mohr}, {Montroy}, {Ngeow}, {Padin}, {Plagge},
  {Pryke}, {Reichardt}, {Rest}, {Ruel}, {Ruhl}, {Schaffer}, {Shirokoff},
  {Song}, {Spieler}, {Stalder}, {Staniszewski}, {Stark}, {Stubbs}, {van
  Engelen}, {Vieira}, {Williamson}, {Yang}, {Zahn}, \&
  {Zenteno}}]{vanderlinde10}
{Vanderlinde} K. {et~al.}, 2010, \apj, 722, 1180

\bibitem[{{Williamson} {et~al}\mbox{.}(2011){Williamson}, {Benson}, {High},
  {Vanderlinde}, {Ade}, {Aird}, {Andersson}, {Armstrong}, {Ashby}, {Bautz},
  {Bazin}, {Bertin}, {Bleem}, {Bonamente}, {Brodwin}, {Carlstrom}, {Chang},
  {Chapman}, {Clocchiatti}, {Crawford}, {Crites}, {de Haan}, {Desai}, {Dobbs},
  {Dudley}, {Fazio}, {Foley}, {Forman}, {Garmire}, {George}, {Gladders},
  {Gonzalez}, {Halverson}, {Holder}, {Holzapfel}, {Hoover}, {Hrubes}, {Jones},
  {Joy}, {Keisler}, {Knox}, {Lee}, {Leitch}, {Lueker}, {Luong-Van}, {Marrone},
  {McMahon}, {Mehl}, {Meyer}, {Mohr}, {Montroy}, {Murray}, {Padin}, {Plagge},
  {Pryke}, {Reichardt}, {Rest}, {Ruel}, {Ruhl}, {Saliwanchik}, {Saro},
  {Schaffer}, {Shaw}, {Shirokoff}, {Song}, {Spieler}, {Stalder}, {Stanford},
  {Staniszewski}, {Stark}, {Story}, {Stubbs}, {Vieira}, {Vikhlinin}, \&
  {Zenteno}}]{williamson11}
{Williamson} R. {et~al.}, 2011, \apj, 738, 139

\end{thebibliography}

\end{document}